\documentclass[useAMS,usenatbib]{mnras}

\title[Criterions for retrograde rotation of accreting black holes]{Criterions for retrograde rotation of accreting black holes}

\author[A.G. Mikhailov et al.]{
A.G. Mikhailov, \thanks{E-mail: mag10629@yandex.ru}
M.Yu. Piotrovich \thanks{E-mail: mpiotrovich@mail.ru},
Yu.N. Gnedin \thanks{E-mail: gnedin@gaoran.ru},
T.M. Natsvlishvili,
\newauthor S.D. Buliga
\\
Central Astronomical Observatory at Pulkovo, 196140, Saint-Petersburg, Russia}

\date{Accepted for publication in MNRAS.}

\pubyear{2018}

\begin{document}

\label{firstpage}
\pagerange{\pageref{firstpage}--\pageref{lastpage}}
\maketitle

\begin{abstract}
Rotating supermassive black holes produce jets and their origin is connected to magnetic field that is generated by accreting matter flow. There is a point of view that electromagnetic fields around rotating black holes are brought to the hole by accretion. In this situation the prograde accreting disks produce weaker large-scale black hole threading magnetic fields, implying weaker jets that in retrograde regimes. The basic goal of this paper is to find the best candidates for retrograde accreting systems in observed active galactic nuclei. We show that active galactic nuclei with low Eddington ratio are really the best candidates for retrograde systems. This conclusion is obtained for kinetically dominated FRII radio galaxies, flat spectrum radio loud  narrow line Seyfert I galaxies and a number of nearby galaxies. Our conclusion is that the best candidates for retrograde systems are the noticeable population of active galactic nuclei in the Universe. This result corresponds to the conclusion that in the merging process the interaction of merging black holes with a retrograde circumbinary disk is considerably more effective for shrinking the binary system.
\end{abstract}


\section{Introduction}%{1}

Active galactic nuclei (AGN) produce relativistic outflows (jets). The general consensus is that accretion disks, magnetic fields and rotating black holes are the most important factors for the generation of the powerful energy in AGNs. Mediated by large scale magnetic fields, relativistic jets can be powered by the central black hole (BH) via Blandford-Znajek \citep{blandford77}, Blandford-Payne \citep{blandford82} and Meier \citep{meier99} mechanisms. The jet power satisfies \citep{daly14} relation:

\begin{equation}
 L_j \sim B_H^2 M_{BH}^2 a^2,
 \label{eq01}
\end{equation}

\noindent where $L_j$ is the beam power of the jet, $M_{BH}$ is the BH mass, $a$ is the dimensionless spin value and $B_H$ is the poloidal magnetic field strength at the event horizon radius of a BH.

According to \citet{daly09,daly11} and \citet{daly14} the relationship (\ref{eq01}) implies that the BH spin $a$ can be determined if the relativistic jet power, BH mass and magnetic field can be estimated. As a result \citet{daly14} obtained the following expression for the spin:

\begin{equation}
 |a| = \eta \left(\frac{L_j}{10^{44} erg/s}\right)^{0.5} \left(\frac{10^4 G}{B_H}\right) \left(\frac{10^8 M_{\odot}}{M_{BH}}\right).
 \label{eq02}
\end{equation}

\noindent Constant $\eta$ depends on the models. For example, for the model of \citet{blandford77} $\eta = \sqrt{5}$, and for the hybrid model of \citet{meier99} $\eta = 1.05^{-1/2}$.

The spin value $a$ is the dimensionless parameter, i.e. $a = c J / G M_{BH}^2$, where $J$ is the angular momentum. The spin can take on values $-0.998 \leq a \leq 0.998$ \citep{thorne74}. The negative values correspond to the situation when the BH rotation is retrograde with respect to the rotation of the accretion disk. The jet efficiency depends on the possibility of the accretion flow to drag large scale magnetic field to the central BH. As a result, the magnetic field at the event horizon radius of BH is produced by accreting flow and therefore may depend on the ratio between energy densities (pressures) of the accretion flow and magnetic field, i.e. on $\beta = P_{acc} / P_{magn}$, where $P_{acc}$ is the accretion flow pressure and $P_{magn}$ is the magnetic field pressure. The case of $\beta = 1.0$ corresponds to the equality of pressures.

There is the interesting problem how the power of the relativistic jet depends on the sign of the spin parameter, i.e. for prograde ($a > 0$) and retrograde ($a < 0$) rotations. This question is widely discussed in many publications \citep{hawley06,mckinney09,garofalo09,sadowski10,komissarov11,garofalo10,tchekhovskoy10,tchekhovskoy12,gnedin13,garofalo13}. Many radiative and magnetohydrodynamics (MHD) processes have been examined as possible jet launch. Jets are traditionally considered as MHD phenomenon, often associated also with an accretion disk. \citet{garofalo09} has developed a scenario in which the central inward ''plunging'' region of the accretion flow enhances the trapping of a large scale poloidal field on the BH. He concluded that magnetic field dragging results produce higher jet efficiency, namely for retrograde black holes. This situation was discussed in detail by \citet{reynolds06}.

\citet{tchekhovskoy12} have studied prograde and retrograde disk accretion on rapidly spinning BHs via 3D time dependent general relativistic MHD simulations. They discovered that the efficiency with which accreting BHs generate energy power of jets depends not only on the BH spin, but also on as accretion disk thickness $H/R$. As a result they found that prograde BHs with thick disk ($H/R = 0.3 - 0.6$) generate jets several times more efficiently than retrograde BHs with the same value of $|a|$.

In this situation the global problem is to formulate the necessary criterions for producing the retrograde accretion system. This ia our basic goal in this paper. We will also demonstrate the examples of AGNs which can be considered as the objects with retrograde supermassive black holes (SMBH). This is our basic goal in this paper.

\section{The basic methods and results}%2

We will use the relation (\ref{eq01}) between the dimensionless BH spin $a$, the jet power $L_j$, BH mass $M_{BH}$ and the magnetic field $B_H$, threading the BH ergosphere and the accretion disk itself \citep{daly14}. For formulating criterions for the retrograde system we use this relation in the hybrid model of \citet{meier99}.

The basic problem is determining the magnetic field strength $B_H$ on the event horizon radius $R_H$ that is determined as \citep{novikov73}:

\begin{equation}
 R_H = \frac{G M_{BH}}{c^2} \left(1 + \sqrt{1 - a^2}\right).
 \label{eq03}
\end{equation}

\noindent The magnetic field $B_H$, that is responsible for generation of the relativistic jet, can be estimated using the magnetic coupling (MC) model, developed by \citet{li02,ma07,wang02,wang03,zhang05}. In this model the magnetic field $B_H$ is produced by the accreting flow, interacting with the rotating BH. As a result, the magnetic field strength of BH can be determined by the relation between the energy densities (pressures) of the accretion flow and the magnetic field \citep{wang02,wang03,ma07,silantev09}:

\[
 B_H = \frac{1}{R_H} \sqrt{\frac{2 \dot{M} c}{\beta}} =
\]
\begin{equation}
  = 6.3 \times 10^4 \left(\frac{l_E}{M_8}\right)^{1/2} \left(\frac{1}{\beta \varepsilon}\right)^{1/2} \frac{1}{1 + \sqrt{1 - a^2}},
 \label{eq04}
\end{equation}

\noindent where $\dot{M}$ is the accretion rate, $l_E = L_{bol} / L_{Edd}$ is the Eddington factor, $M_8 = M_{BH} / 10^8 M_{\odot}$ and $\beta = P_{acc} / P_{magn}$ is the ratio between the energy densities of the accretion flow and magnetic field. The equipartition corresponds to the situation when $\beta = 1.0$. $\varepsilon$ is the coefficient of radiation efficiency of the accretion flow. This coefficient depends strongly on the BH spin \citep{krolik07,hawley07}.

We can estimate the value of the BH spin using the equations (\ref{eq01})-(\ref{eq04}) which are transformed in the following basic form:

\begin{equation}
 f(a) = \frac{|a|}{\sqrt{\varepsilon (a)}(1 + \sqrt{1 - a^2})} = 1.77 \times \sqrt{\beta} \left(\frac{L_j}{L_{bol}}\right)^{1/2}.
 \label{eq05}
\end{equation}

We use the hybrid model of \citet{meier99}, that describes the process of transformation of the accretion flow energy into the generation of the magnetic field $B_H$. In this situation $\beta \geq 1.0$.

For the retrograde rotation ($a < 0$) the maximal value of $f_{max} = f(-0.998) = 4.8$. For $a = 0.998$ the value is $f(0.998) = 1.66$. The dependence of the function of the spin $a$ is presented on the Fig.1 from \citet{gnedin14}.

Now we start with solving the basic problem - how it is possible to get the observational criterions for determining the retrograde systems.

There is the fundamental plane of BH activity that indicates a relationship between compact radio emission, X-ray luminosity and BH mass. This plane may be considered as a manifestation of the real relationship between jet power $L_j$, bolometric accretion disk luminosity and BH mass (or Eddington luminosity $L_{Edd}$). \citet{merloni03} studied the correlation between the radio ($L_R$) and the X-ray ($L_X$) luminosities and the BH mass ($M_{BH}$). They showed that these sources define ''Fundamental Plane'' for SMBHs in the three-dimensional ($\log{L_R}, \log{L_X}, \log{M_{BH}}$) space. They confirmed that the models of radiatively inefficient accretion flows, which have been used in Eq.(\ref{eq05}), are in quite well agreement with observed data. \citet{merloni07} and \citet{daly16} obtained the following dimensionless form:

\begin{equation}
 \log{\frac{L_j}{L_{Edd}}} = A \log{l_E} + B.
 \label{eq06}
\end{equation}

\noindent The real estimates of $A$ and $B$ are presented in their papers. Thus according to \citet{merloni07} we have $A = 0.49\pm 0.07$ and $B = - (0.78 \pm 0.36)$. \citet{daly16} obtained via observations of 80 AGNs and 36 X-ray binaries $A = 0.45 \pm 0.05$, $B = -0.94 \pm 0.16$ and from results of observations of 80 AGNs $A = 0.41 \pm 0.04$ and $B = -1.34 \pm 0.14$.

As a result we obtain the following relation for estimating the spin value from Eq.(\ref{eq05}) and (\ref{eq06}) using \citet{merloni07} data:

\begin{equation}
 \left(\frac{L_j}{L_{bol}}\right)^{1/2} = 10^{B/2} l_E^{(A - 1)/2},
 \label{eq07}
\end{equation}

\begin{equation}
 f(a) = \sqrt{\beta} \times 1.77 \times 10^{-(0.39 \pm 0.18)} l_E^{-(0.255 \pm 0.035)}.
 \label{eq08}
\end{equation}

\citet{daly16} have obtained the values of the parameters $A$ and $B$ for nine samples of AGNs and X-ray binaries, including LINERS.

\begin{equation}
 f(a) = \sqrt{\beta} \times 1.77 \times 10^{-(0.47 \pm 0.05)} l_E^{-(0.275 \pm 0.015)},
 \label{eq09}
\end{equation}

\begin{equation}
 f(a) = \sqrt{\beta} \times 1.77 \times 10^{-(0.67 \pm 0.07)} l_E^{-(0.295 \pm 0.02)},
 \label{eq10}
\end{equation}

\noindent The expressions (\ref{eq07}) - (\ref{eq10}) are obtained for \citet{meier99} model. For BZ mechanism we have $f(BZ) = \sqrt{5.25} f(Meier)$.

The main problem is existing noticeable interval between values $a > 0$ and $a < 0$ where values $f(a)$ and $f(-a)$ are strongly intersected with each other. For example, the values $f(0.7) \approx f (-0.5)$. Real deriving of the retrograde movement takes place at $a < -0.6$. In this case $f(a) > 1.7$. The maximal value for positive $a$ is $f(0.998) = 1.66$. Thus the boundary between prograde and retrograde rotations is $f = 1.7$.

According to \citet{garofalo13} in the case when the accretion is the only physical process providing the generation of the magnetic field in the core of the relativistic jet the parameter $\beta$ can not be less than 1. This situation permits to get the criterion on the Eddington ratio $l_E$ in the case of the retrograde rotation. For example, for \citet{merloni07} relation the limiting value for function $f$ is $f(a) = 1.77 \sqrt{\beta} \times 10^{-0.57} l_E^{-0.22}$ and for pure retrograde rotation ($a = -0.625$) we obtained $l_E = 10^{-2.5}$. Thus the retrograde objects are AGNs with the values of the Eddington ratio $l_E \leq 10^{-2.5}$.

In the modern publications enough AGN data are presented with $L_E \leq 10^{-2.5}$. This value is valid for hybrid model of the jet generation \citep{meier99}. For BZ mechanism the situation is more exciting. In this case one needs the additional coefficient in Eq.(\ref{eq05}), that is equal to $\sqrt{5.25}$. As a result one obtains the required condition for retrograde regime: $l_E \leq 0.1$, only for BZ process.

Obtained results allow us to derive the best candidates for the retrograde SMBHs.

We have analyzed in detail the FRII objects. The question arises about the origin of FRII objects. Our estimates show that the retrograde systems exist also in FRI objects. The list of the retrograde FRI radio galaxies included B2~0755+137, 3C~31, 3C~317, B2~0055+30, 3C~66B, 3C~449, 3C~2721 from the sample presented in \citet{wu09}.

\section{Kinetically dominated FRII radio objects}%3

According to (\ref{eq05}), one can expect the retrograde rotation in the objects where the jet kinetic power $L_j$ exceeds essentially the bolometric luminosity $L_{bol}$. For example, most quasars are radio-quiet, but most of Fanaroff-Riley class II (FRII) quasars have powerful radio lobes \citep{punsly07}. Estimates of these radio lobes showed that the value $L_j$ is enormous essentially higher than the bolometric luminosity $L_{bol}$.

Let us consider as an example a powerful, kinetically dominated quasar PKS~1018-42 \citep{punsly06}. This object is a radio galaxy with extraordinarily powerful jet with kinetic luminosity $L_j = 6.5 \times 10^{46} erg/s$. This value is 3.4 times larger that the total luminosity of the accretion flow $L_{bol} = 1.9 \times 10^{46} erg/s$. As a result the solution of (\ref{eq05}) gives $f(a) = 3.27 \sqrt{\beta}$. It means that for the acting accretion process with $\beta \geq 1.0$ the value of the spin is $a \leq -0.91$, i.e. we have the retrograde candidate.

\begin{table}
\begin{center}
\caption{Kinetically dominated FRII radio galaxies - candidates in retrograde systems \citep{punsly06,punsly07,wu09}}
\begin{tabular}{lcccc}
\hline
Object      & $\log{L_{bol}}$ & $\log{L_j}$ & $f(a)$      & $a$ \\
            &                 &             & $\beta = 1$ & up. limit \\
\hline
PKS 1018-42 & 46.28           & 46.814      & 3.27        & -0.91 \\
3C 190      & 45.85           & 46.63       & 4.34        & -0.96 \\
3C 455      & 44.99           & 45.84       & 4.71        & -0.95 \\
3C 82       & 46.4            & 47.2        & 4.45        & -0.95 \\
3C 9        & 46.59           & 47.17       & 3.45        & -0.92 \\
4C 25.21    & 46.06           & 46.77       & 4.00        & -0.93 \\
4C 04.81    & 46.55           & 47.017      & 3.04        & -0.89 \\
3C 196      & 46.50           & 46.87       & 2.94        & -0.85 \\
3C 14       & 46.51           & 46.71       & 2.28        & -0.75 \\
3C 270.1    & 46.68           & 46.81       & 2.08        & -0.71 \\
\hline
\end{tabular}
\label{tab1}
\end{center}
\end{table}

The results for other objects from \citet{punsly06,punsly07,wu09} are presented at the Table \ref{tab1}.

The object 3C~216 deserves the special attention. According to \citet{punsly07} the physical parameters of this object are the following: $l_E = 0.05 - 1.0$ and $L_j / L_{Edd} = 3.3 - 10$. The value of the function $f(a)$ is $f(a) = 10.17 \sqrt{\beta}$. It means that the maximal retrograde orbit ($a = -1$) can be reached only if $\beta = 0.256$, that does not correspond to the traditional accretion model. This situation requires the effective mechanism of magnetic field generation, that is not connected directly to the accretion process. Of course, this situation requires the special investigation.

\section{Retrograde SMBHs in the sample of FRII objects with power radio fluxes}%4

\begin{table}
\begin{center}
\caption{The sample of FRII radio galaxies with measured power radio fluxes \citep{chiaberge02}}
\begin{tabular}{lcccc}
\hline
Object      & $\log{L_{bol}}$ & $\log{L_j}$ & $f(a)$      & $a$ \\
            &                 &             & $\beta = 1$ & up. limit \\
\hline

3C 17&
45.45&
46.29&
4.65&
-0.995 \\

3C 18&
45.42&
45.55&
2.04&
-0.71 \\

3C 35&
43.57&
44.28&
3.97&
-0.97 \\

3C 105&
44.32&
44.35&
1.82&
-0.66 \\

3C 123&
45.12&
45.53&
2.83&
-0.85 \\

3C 173.1&
44.89&
45.01&
2.03&
-0.71 \\

3C 219&
45.16&
45.30&
2.08&
-0.72 \\

3C 326&
44.18&
44.33&
2.10&
-0.72 \\

3C 349&
44.96&
45.01&
1.87&
-0.67 \\

3C 388&
44.03&
44.90&
4.79&
-0.997 \\

3C 401&
44.55&
45.27&
4.05&
-0.97 \\

3C 436&
45.06&
45.04&
1.72&
-0.63 \\

3C 460&
45.21&
45.18&
1.71&
-0.63 \\
\hline
\end{tabular}
\label{tab2}
\end{center}
\end{table}

\begin{table*}
\begin{center}
\caption{Retrograde SMBHs in flat spectrum radio loud narrow line Seyfert 1 galaxies \citep{berton16}}
\begin{tabular}{lcccccc}
\hline
Object & $\log{\frac{M_{BH}}{M_{\odot}}}$  & $\log{L_{bol}}$ & $\log{l_{E}}$ & $\log{L_{j}}$ & $f(a)$      & $a$ \\
       &                 &                 &               &               & $\beta = 1$ & up. limit \\
\hline
J0022833.42+005510.9 & 8.94 & 44.10 & -2.96 & 44.83 & $4.10_{-0.66}^{+0.79}$ & $-0.98$ \\
J075756.71+395936.0  & 7.13 & 43.77 & -1.52 & 43.74 & $1.76_{-0.45}^{+0.60}$ & $-0.64$ \\
J114311.01+053516.1  & 8.84 & 45.08 & -2.00 & 45.26	& $2.33_{-0.52}^{+0.67}$ & $-0.77$ \\
J140416.35+411748.7  & 7.96 & 43.88 & -2.22 & 44.22	& $2.65_{-0.55}^{+0.71}$ & $-0.83$ \\
J140942.44+360415.8  & 8.24 & 43.82 & -2.52 & 44.33	& $3.17_{-0.61}^{+0.74}$ & $-0.90$ \\
\hline
\end{tabular}
\label{tab3}
\end{center}
\end{table*}

\begin{table}
\begin{center}
\caption{The basic characteristics of SMBHs in nearby galaxies \citep{ho99}}
\begin{tabular}{lcccc}
\hline
Object   & $L_{bol}$            & $l_E$                & $L_j$         & $a$ \\
         & [erg/s]              &                      & [erg/s]       & up. limit \\
\hline
NGC 3031 & $2.1 \times 10^{41}$ & $4.2 \times 10^{-4}$ & $10^{41.74} $ & -0.88 \\
NGC 4261 & $1.7 \times 10^{42}$ & $2.8 \times 10^{-5}$ & $10^{43.28} $ & -1.0 \\
NGC 4374 & $8.2 \times 10^{41}$ & $4.3 \times 10^{-6}$ & $10^{42.50} $ & -0.92 \\
NGC 4486 & $2.3 \times 10^{42}$ & $6.1 \times 10^{-6}$ & $10^{43.11} $ & -0.98 \\
NGC 4579 & $9.9 \times 10^{41}$ & $1.9 \times 10^{-3}$ & $10^{42.17} $ & -0.97 \\
NGC 4594 & $2.7 \times 10^{41}$ & $2.2 \times 10^{-6}$ & $10^{42.50} $ & -0.95 \\
NGC 6251 & $8.0 \times 10^{42}$ & $8.5 \times 10^{-5}$ & $10^{44.50} $ & -0.84 \\
\hline
\end{tabular}
\label{tab4}
\end{center}
\end{table}

\citet{chiaberge02} have studied the nuclei of 3CR FRII galaxies through HST optical images up to $z = 0.3$. They analyzed these objects determining their radio and optical nuclear luminosities. They derived the radio core luminosity at 5GHz and nuclear luminosity in OIII wavelength. These data allows us to estimate the SMBH spin using Eq.(\ref{eq05}). The bolometric luminosity can be estimated via the fundamental relation $L_{bol} = 3500 L(OIII)$ and the kinetic power of a jet via the following relation $L_j = 10^{11.9} L_R^{0.81}$ \citep{merloni07}. Thus we distinguished the retrograde systems which are presented at the Table \ref{tab2}. For a number of 3C objects we obtained also the extremely high value of the parameter $f(a)$. These objects included 3C~15, 3C~40, 3C~88, 3C~236, 3C~353.

\section{Retrograde SMBHs in flat spectrum radio loud narrow line Seyfert 1 galaxies}%5

Narrow line Seyfert 1 galaxies are an special subclass of AGN. Many of them does not exhibit any strong radio emission. But many of them are the sources of $\gamma$-ray emission. This fact is usually interpreted as a sign of relativistic beamed jet oriented along the line of sight. \citet{berton16} derived the BH mass and Eddington ratio distributions for these objects. It is interesting that the absence of strong radio emission and presence of essential $\gamma$-ray emission confirmed that the generation of jet kinetic energy is produced by the mechanism described by Eq.(\ref{eq06}) \citep{merloni07,daly16}. For many of these objects $\log{l_E} < -1$ (Table 1 from \citet{berton16}) and this fact allows us to consider these objects as candidates to retrograde systems. This candidates are presented at the Table \ref{tab3}.

As concerns to the Table \ref{tab3} only objects with retrograde systems are presented in this Table. The total quantity of considered AGNs, presented in \citet{berton16}, is quite large and many of them have prograde rotation. All of these objects are narrow-line Seyfert 1 galaxies. They are an interesting subclass of AGNs, which typically does not exhibit any strong radio emission.

\section{Retrograde SMBHs in nearby galaxies}%6

A lot of nearby galaxies are low luminosity AGNs. \citet{ho99} underlined that the spectral energy distribution of AGN carry important information on the physical processes of the accretion mechanism. Many aspects of the AGN phenomenon, including the spectral energy distribution, have been successfully interpreted within the standard accretion disk \citep{shakura73} framework. Unlike the high luminosity AGNs. the low luminosity AGNs can be retrograde systems.

\citet{ho99} presented spectral energy distributions of a sample of seven low luminosity AGNs. The nuclear fluxes were carefully selected to avoid contamination by emission from the host galaxy, which can be essential for very weak nuclei. \citet{ho99} confirmed that the spectral energy distributions (SED) of low luminosity AGNs look markedly different compared to the SEDs in the standard classical luminous AGNs. These peculiarities can be connected with departures in these objects from the standard AGN accretion disk model. We will show that these departures can be explained by the situation that the accretion disk matter in the low luminous AGN is in the retrograde rotation compared to basic SMBH.  It is very important that the Eddington ratio for the objects, presented at the Table 9 of \citet{ho99}, is in the interval $10^{-5.657} \leq l_E \leq 10^{-2.72}$. It means that according to Eqs.(\ref{eq08})-(\ref{eq10}) these AGNs are really the retrograde systems. It is important that at the Table 9 \citep{ho99} the radio luminosity $L_R$ of these objects at $\lambda = 5 \times 10^5$ Hz is also presented. This fact allows us to estimate the spin of SMBHs using the other expression for the jet kinetic power, presented by \citet{merloni07}:

\begin{equation}
 \log{L_j} = (0.81 \pm 0.11) \log{L_R} + 11.9_{-4.4}^{4.1}.
 \label{eq11}
\end{equation}

\noindent Using this relationship we obtained also the negative values of the SMBH spin for all objects from the sample of \citet{ho99}. This result confirms also the retrograde regime for these objects.

The results of determining the spin value for these objects are presented at the Table \ref{tab4}.

\section{Conclusions}%7

AGNs produce power relativistic outflows (jets). Its physical origin is now unclear and requires the detail investigation. Two popular physical mechanisms of jet generation, namely, Blandford-Znajek \citep{blandford77} and Meier \citep{meier99} are now commonly accepted. The difference between these mechanisms is connected with the contribution of the accretion disk. It is appeared that this contribution is dependent on the nature of the rotation of accreting matter in the accretion disk and SMBH itself.

According to \citet{king05}, alignment of BH spin with the angular momentum of an outer accretion disk is produced on short time scale only for thin accretion disk. For large time scale accretion occurs via geometrically thick disk alignment. Also the same situation can take place in the case of warped accretion disk \citep{tremaine14}.

The problem of retrograde versus prograde models of accreting BHs have been recently discussed by \citet{garofalo13,garofalo17}. The commonly accepted point of view is that accretion disks, magnetic fields and rotating BHs are the important elements needed to produce the most powerful sources of energy in the Universe. Rotating BHs produce jets and their origin is connected with electromagnetism. This situation requires to use numerical general relativistic simulations. But it is possible to consider the suggestion that all electromagnetic fields around rotating BHs are brought to the BHs by accretion and only by accretion. In this situation one should await that the magnetic fields near a BH are produced by the accreting matter and then the situation is appeared that the magnetic field energy density is only a part of total energy density of accreting matter, i.e. the parameter $\beta \geq 1$. In this situation the retrograde accretion disks will produce the extremely powerful relativistic jets with $L_j \gg L_{bol}$.

We used \citet{merloni07} data for estimation of a spin value from Eq.(\ref{eq08}). Our estimates showed that AGNs with the Eddington ratio $l_E \leq 10^{-2.5}$ are the retrograde systems. As a result the retrograde regime objects are found in kinetically dominated FRII radio galaxies with powerful radio fluxes in flat spectrum radio loud narrow line Seyfert 1 galaxies and also in a number of nearby galaxies. Our conclusion is that there is a noticeable population of retrograde AGNs in the Universe.

\section*{Acknowledgements}

This research was supported by the Program of Praesidium of Russian Academy of Sciences No.28.

\bibliographystyle{mnras}
\bibliography{mybibfile}

\begin{thebibliography}{}
\makeatletter
\relax
\def\mn@urlcharsother{\let\do\@makeother \do\$\do\&\do\#\do\^\do\_\do\%\do\~}
\def\mn@doi{\begingroup\mn@urlcharsother \@ifnextchar [ {\mn@doi@}
  {\mn@doi@[]}}
\def\mn@doi@[#1]#2{\def\@tempa{#1}\ifx\@tempa\@empty \href
  {http://dx.doi.org/#2} {doi:#2}\else \href {http://dx.doi.org/#2} {#1}\fi
  \endgroup}
\def\mn@eprint#1#2{\mn@eprint@#1:#2::\@nil}
\def\mn@eprint@arXiv#1{\href {http://arxiv.org/abs/#1} {{\tt arXiv:#1}}}
\def\mn@eprint@dblp#1{\href {http://dblp.uni-trier.de/rec/bibtex/#1.xml}
  {dblp:#1}}
\def\mn@eprint@#1:#2:#3:#4\@nil{\def\@tempa {#1}\def\@tempb {#2}\def\@tempc
  {#3}\ifx \@tempc \@empty \let \@tempc \@tempb \let \@tempb \@tempa \fi \ifx
  \@tempb \@empty \def\@tempb {arXiv}\fi \@ifundefined
  {mn@eprint@\@tempb}{\@tempb:\@tempc}{\expandafter \expandafter \csname
  mn@eprint@\@tempb\endcsname \expandafter{\@tempc}}}

\bibitem[\protect\citeauthoryear{{Berton} et~al.,}{{Berton}
  et~al.}{2016}]{berton16}
{Berton} M.,  et~al., 2016, \mn@doi [\aap] {10.1051/0004-6361/201628171}, \href
  {http://adsabs.harvard.edu/abs/2016A%26A...591A..98B} {591, A98}

\bibitem[\protect\citeauthoryear{{Blandford} \& {Payne}}{{Blandford} \&
  {Payne}}{1982}]{blandford82}
{Blandford} R.~D.,  {Payne} D.~G.,  1982, \mn@doi [\mnras]
  {10.1093/mnras/199.4.883}, \href
  {http://adsabs.harvard.edu/abs/1982MNRAS.199..883B} {199, 883}

\bibitem[\protect\citeauthoryear{{Blandford} \& {Znajek}}{{Blandford} \&
  {Znajek}}{1977}]{blandford77}
{Blandford} R.~D.,  {Znajek} R.~L.,  1977, \mn@doi [\mnras]
  {10.1093/mnras/179.3.433}, \href
  {http://adsabs.harvard.edu/abs/1977MNRAS.179..433B} {179, 433}

\bibitem[\protect\citeauthoryear{{Chiaberge}, {Capetti}  \&
  {Celotti}}{{Chiaberge} et~al.}{2002}]{chiaberge02}
{Chiaberge} M.,  {Capetti} A.,   {Celotti} A.,  2002, \mn@doi [\aap]
  {10.1051/0004-6361:20021204}, \href
  {http://adsabs.harvard.edu/abs/2002A%26A...394..791C} {394, 791}

\bibitem[\protect\citeauthoryear{{Daly}}{{Daly}}{2009}]{daly09}
{Daly} R.~A.,  2009, \mn@doi [\apjl] {10.1088/0004-637X/696/1/L32}, \href
  {http://adsabs.harvard.edu/abs/2009ApJ...696L..32D} {696, L32}

\bibitem[\protect\citeauthoryear{{Daly}}{{Daly}}{2011}]{daly11}
{Daly} R.~A.,  2011, \mn@doi [\mnras] {10.1111/j.1365-2966.2011.18452.x}, \href
  {http://adsabs.harvard.edu/abs/2011MNRAS.414.1253D} {414, 1253}

\bibitem[\protect\citeauthoryear{{Daly} \& {Sprinkle}}{{Daly} \&
  {Sprinkle}}{2014}]{daly14}
{Daly} R.~A.,  {Sprinkle} T.~B.,  2014, \mn@doi [\mnras]
  {10.1093/mnras/stt2433}, \href
  {http://adsabs.harvard.edu/abs/2014MNRAS.438.3233D} {438, 3233}

\bibitem[\protect\citeauthoryear{{Daly}, {Stout}  \& {Mysliwiec}}{{Daly}
  et~al.}{2016}]{daly16}
{Daly} R.~A.,  {Stout} D.~A.,   {Mysliwiec} J.~N.,  2016, preprint, \href
  {http://adsabs.harvard.edu/abs/2016arXiv160601399D} {} (\mn@eprint {arXiv}
  {1606.01399})

\bibitem[\protect\citeauthoryear{{Garofalo}}{{Garofalo}}{2009}]{garofalo09}
{Garofalo} D.,  2009, \mn@doi [\apj] {10.1088/0004-637X/699/1/400}, \href
  {http://adsabs.harvard.edu/abs/2009ApJ...699..400G} {699, 400}

\bibitem[\protect\citeauthoryear{{Garofalo}}{{Garofalo}}{2013}]{garofalo13}
{Garofalo} D.,  2013, \mn@doi [\mnras] {10.1093/mnras/stt1237}, \href
  {http://adsabs.harvard.edu/abs/2013MNRAS.434.3196G} {434, 3196}

\bibitem[\protect\citeauthoryear{{Garofalo}}{{Garofalo}}{2017}]{garofalo17}
{Garofalo} D.,  2017, \mn@doi [\apss] {10.1007/s10509-017-3108-x}, \href
  {http://adsabs.harvard.edu/abs/2017Ap%26SS.362..121G} {362, 121}

\bibitem[\protect\citeauthoryear{{Garofalo}, {Evans}  \& {Sambruna}}{{Garofalo}
  et~al.}{2010}]{garofalo10}
{Garofalo} D.,  {Evans} D.~A.,   {Sambruna} R.~M.,  2010, \mn@doi [\mnras]
  {10.1111/j.1365-2966.2010.16797.x}, \href
  {http://adsabs.harvard.edu/abs/2010MNRAS.406..975G} {406, 975}

\bibitem[\protect\citeauthoryear{{Gnedin}, {Piotrovich}, {Buliga}  \&
  {Natsvlishvili}}{{Gnedin} et~al.}{2013}]{gnedin13}
{Gnedin} Y.~N.,  {Piotrovich} M.~Y.,  {Buliga} S.~D.,   {Natsvlishvili} T.~M.,
  2013, \mn@doi [Astronomische Nachrichten] {10.1002/asna.201211844}, \href
  {http://adsabs.harvard.edu/abs/2013AN....334..264G} {334, 264}

\bibitem[\protect\citeauthoryear{{Gnedin}, {Globina}, {Piotrovich}, {Buliga}
  \& {Natsvlishvili}}{{Gnedin} et~al.}{2014}]{gnedin14}
{Gnedin} Y.~N.,  {Globina} V.~N.,  {Piotrovich} M.~Y.,  {Buliga} S.~D.,
  {Natsvlishvili} T.~M.,  2014, \mn@doi [Astrophysics]
  {10.1007/s10511-014-9323-z}, \href
  {http://adsabs.harvard.edu/abs/2014Ap.....57..163G} {57, 163}

\bibitem[\protect\citeauthoryear{{Hawley} \& {Krolik}}{{Hawley} \&
  {Krolik}}{2006}]{hawley06}
{Hawley} J.~F.,  {Krolik} J.~H.,  2006, \mn@doi [\apj] {10.1086/500385}, \href
  {http://adsabs.harvard.edu/abs/2006ApJ...641..103H} {641, 103}

\bibitem[\protect\citeauthoryear{{Hawley}, {Beckwith}  \& {Krolik}}{{Hawley}
  et~al.}{2007}]{hawley07}
{Hawley} J.~F.,  {Beckwith} K.,   {Krolik} J.~H.,  2007, \mn@doi [\apss]
  {10.1007/s10509-007-9559-8}, \href
  {http://adsabs.harvard.edu/abs/2007Ap%26SS.311..117H} {311, 117}

\bibitem[\protect\citeauthoryear{{Ho}}{{Ho}}{1999}]{ho99}
{Ho} L.~C.,  1999, \mn@doi [\apj] {10.1086/307137}, \href
  {http://adsabs.harvard.edu/abs/1999ApJ...516..672H} {516, 672}

\bibitem[\protect\citeauthoryear{{King}, {Lubow}, {Ogilvie}  \&
  {Pringle}}{{King} et~al.}{2005}]{king05}
{King} A.~R.,  {Lubow} S.~H.,  {Ogilvie} G.~I.,   {Pringle} J.~E.,  2005,
  \mn@doi [\mnras] {10.1111/j.1365-2966.2005.09378.x}, \href
  {http://adsabs.harvard.edu/abs/2005MNRAS.363...49K} {363, 49}

\bibitem[\protect\citeauthoryear{{Komissarov}}{{Komissarov}}{2011}]{komissarov11}
{Komissarov} S.~S.,  2011, \memsai, \href
  {http://adsabs.harvard.edu/abs/2011MmSAI..82...95K} {82, 95}

\bibitem[\protect\citeauthoryear{{Krolik}}{{Krolik}}{2007}]{krolik07}
{Krolik} J.~H.,  2007, preprint, \href
  {http://adsabs.harvard.edu/abs/2007arXiv0709.1489K} {} (\mn@eprint {arXiv}
  {0709.1489})

\bibitem[\protect\citeauthoryear{{Li}}{{Li}}{2002}]{li02}
{Li} L.-X.,  2002, \mn@doi [\apj] {10.1086/338486}, \href
  {http://adsabs.harvard.edu/abs/2002ApJ...567..463L} {567, 463}

\bibitem[\protect\citeauthoryear{{Ma}, {Yuan}  \& {Wang}}{{Ma}
  et~al.}{2007}]{ma07}
{Ma} R.-Y.,  {Yuan} F.,   {Wang} D.-X.,  2007, \mn@doi [\apj] {10.1086/522917},
  \href {http://adsabs.harvard.edu/abs/2007ApJ...671.1981M} {671, 1981}

\bibitem[\protect\citeauthoryear{{McKinney} \& {Blandford}}{{McKinney} \&
  {Blandford}}{2009}]{mckinney09}
{McKinney} J.~C.,  {Blandford} R.~D.,  2009, \mn@doi [\mnras]
  {10.1111/j.1745-3933.2009.00625.x}, \href
  {http://adsabs.harvard.edu/abs/2009MNRAS.394L.126M} {394, L126}

\bibitem[\protect\citeauthoryear{{Meier}}{{Meier}}{1999}]{meier99}
{Meier} D.~L.,  1999, \mn@doi [\apj] {10.1086/307671}, \href
  {http://adsabs.harvard.edu/abs/1999ApJ...522..753M} {522, 753}

\bibitem[\protect\citeauthoryear{{Merloni} \& {Heinz}}{{Merloni} \&
  {Heinz}}{2007}]{merloni07}
{Merloni} A.,  {Heinz} S.,  2007, \mn@doi [\mnras]
  {10.1111/j.1365-2966.2007.12253.x}, \href
  {http://adsabs.harvard.edu/abs/2007MNRAS.381..589M} {381, 589}

\bibitem[\protect\citeauthoryear{{Merloni}, {Heinz}  \& {di Matteo}}{{Merloni}
  et~al.}{2003}]{merloni03}
{Merloni} A.,  {Heinz} S.,   {di Matteo} T.,  2003, \mn@doi [\mnras]
  {10.1046/j.1365-2966.2003.07017.x}, \href
  {http://adsabs.harvard.edu/abs/2003MNRAS.345.1057M} {345, 1057}

\bibitem[\protect\citeauthoryear{{Novikov} \& {Thorne}}{{Novikov} \&
  {Thorne}}{1973}]{novikov73}
{Novikov} I.~D.,  {Thorne} K.~S.,  1973, in {Dewitt} C.,  {Dewitt} B.~S.,  eds,
  Black Holes (Les Astres Occlus). pp 343--450

\bibitem[\protect\citeauthoryear{{Punsly}}{{Punsly}}{2007}]{punsly07}
{Punsly} B.,  2007, \mn@doi [\mnras] {10.1111/j.1745-3933.2006.00254.x}, \href
  {http://adsabs.harvard.edu/abs/2007MNRAS.374L..10P} {374, L10}

\bibitem[\protect\citeauthoryear{{Punsly} \& {Tingay}}{{Punsly} \&
  {Tingay}}{2006}]{punsly06}
{Punsly} B.,  {Tingay} S.~J.,  2006, \mn@doi [\apjl] {10.1086/503277}, \href
  {http://adsabs.harvard.edu/abs/2006ApJ...640L..21P} {640, L21}

\bibitem[\protect\citeauthoryear{{Reynolds}, {Garofalo}  \&
  {Begelman}}{{Reynolds} et~al.}{2006}]{reynolds06}
{Reynolds} C.~S.,  {Garofalo} D.,   {Begelman} M.~C.,  2006, \mn@doi [\apj]
  {10.1086/507691}, \href {http://adsabs.harvard.edu/abs/2006ApJ...651.1023R}
  {651, 1023}

\bibitem[\protect\citeauthoryear{{S{\c a}dowski} \& {Sikora}}{{S{\c a}dowski}
  \& {Sikora}}{2010}]{sadowski10}
{S{\c a}dowski} A.,  {Sikora} M.,  2010, \mn@doi [\aap]
  {10.1051/0004-6361/201014076}, \href
  {http://adsabs.harvard.edu/abs/2010A%26A...517A..18S} {517, A18}

\bibitem[\protect\citeauthoryear{{Shakura} \& {Sunyaev}}{{Shakura} \&
  {Sunyaev}}{1973}]{shakura73}
{Shakura} N.~I.,  {Sunyaev} R.~A.,  1973, \aap, \href
  {http://adsabs.harvard.edu/abs/1973A%26A....24..337S} {24, 337}

\bibitem[\protect\citeauthoryear{{Silant'ev}, {Piotrovich}, {Gnedin}  \&
  {Natsvlishvili}}{{Silant'ev} et~al.}{2009}]{silantev09}
{Silant'ev} N.~A.,  {Piotrovich} M.~Y.,  {Gnedin} Y.~N.,   {Natsvlishvili}
  T.~M.,  2009, \mn@doi [\aap] {10.1051/0004-6361/200810892}, \href
  {http://adsabs.harvard.edu/abs/2009A%26A...507..171S} {507, 171}

\bibitem[\protect\citeauthoryear{{Tchekhovskoy} \& {McKinney}}{{Tchekhovskoy}
  \& {McKinney}}{2012}]{tchekhovskoy12}
{Tchekhovskoy} A.,  {McKinney} J.~C.,  2012, \mn@doi [\mnras]
  {10.1111/j.1745-3933.2012.01256.x}, \href
  {http://adsabs.harvard.edu/abs/2012MNRAS.423L..55T} {423, L55}

\bibitem[\protect\citeauthoryear{{Tchekhovskoy}, {Narayan}  \&
  {McKinney}}{{Tchekhovskoy} et~al.}{2010}]{tchekhovskoy10}
{Tchekhovskoy} A.,  {Narayan} R.,   {McKinney} J.~C.,  2010, \mn@doi [\apj]
  {10.1088/0004-637X/711/1/50}, \href
  {http://adsabs.harvard.edu/abs/2010ApJ...711...50T} {711, 50}

\bibitem[\protect\citeauthoryear{{Thorne}}{{Thorne}}{1974}]{thorne74}
{Thorne} K.~S.,  1974, \mn@doi [\apj] {10.1086/152991}, \href
  {http://adsabs.harvard.edu/abs/1974ApJ...191..507T} {191, 507}

\bibitem[\protect\citeauthoryear{{Tremaine} \& {Davis}}{{Tremaine} \&
  {Davis}}{2014}]{tremaine14}
{Tremaine} S.,  {Davis} S.~W.,  2014, \mn@doi [\mnras] {10.1093/mnras/stu663},
  \href {http://adsabs.harvard.edu/abs/2014MNRAS.441.1408T} {441, 1408}

\bibitem[\protect\citeauthoryear{{Wang}, {Xiao}  \& {Lei}}{{Wang}
  et~al.}{2002}]{wang02}
{Wang} D.~X.,  {Xiao} K.,   {Lei} W.~H.,  2002, \mn@doi [\mnras]
  {10.1046/j.1365-8711.2002.05652.x}, \href
  {http://adsabs.harvard.edu/abs/2002MNRAS.335..655W} {335, 655}

\bibitem[\protect\citeauthoryear{{Wang}, {Ma}, {Lei}  \& {Yao}}{{Wang}
  et~al.}{2003}]{wang03}
{Wang} D.-X.,  {Ma} R.-Y.,  {Lei} W.-H.,   {Yao} G.-Z.,  2003, \mn@doi [\apj]
  {10.1086/377303}, \href {http://adsabs.harvard.edu/abs/2003ApJ...595..109W}
  {595, 109}

\bibitem[\protect\citeauthoryear{{Wu}}{{Wu}}{2009}]{wu09}
{Wu} Q.,  2009, \mn@doi [\mnras] {10.1111/j.1365-2966.2009.15127.x}, \href
  {http://adsabs.harvard.edu/abs/2009MNRAS.398.1905W} {398, 1905}

\bibitem[\protect\citeauthoryear{{Zhang}, {Lu}  \& {Zhang}}{{Zhang}
  et~al.}{2005}]{zhang05}
{Zhang} W.-M.,  {Lu} Y.,   {Zhang} S.-N.,  2005, Chinese Journal of Astronomy
  and Astrophysics Supplement, \href
  {http://adsabs.harvard.edu/abs/2005ChJAS...5..347Z} {5, 347}

\makeatother
\end{thebibliography}

\bsp
\label{lastpage}
\end{document}